\documentclass[preprint]{aastex}

\shorttitle{Black Holes and Galactic Bulges} 
\shortauthors{F. C. Adams}

\newcommand{\be}{\begin{equation}}
\newcommand{\ee}{\end{equation}}
 
\newcommand{\mbh}{{ M_{\rm bh}}} 

\newcommand{\tstar}{{ \tau_\ast }} 
\newcommand{\tblg}{{ \tau_{\rm blg} }} 
\newcommand{\tbh}{{ \tau_{\rm bh} }} 
\newcommand{\tcross}{{t_{\rm cross}}}
\newcommand{\fa}{{ {\cal F}_A }} 
\newcommand{\rsw}{ R_S } 
\newcommand{\mrat}{{ \mu_B }} 
\newcommand{\aeff}{ a_{\rm eff} } 
\def\kms{\ifmmode \hbox{ \rm km s}^{-1} \else{ km s$^{-1} $}\fi}

\def\half{{1 \over 2}}
\def\myr{\ifmmode \hbox{\rm Myr}\else{Myr}\fi} 
\def\gyr{\ifmmode \hbox{\rm Gyr}\else{Gyr}\fi} 
\def\kpc{\ifmmode \hbox{\rm  kpc}\else{kpc}\fi} 
\def\Msun{M_\odot}

\begin{document}

\title{A Theoretical Model for the $\mbh-\sigma$ Relation \\
for Supermassive Black Holes in Galaxies} 

\author{Fred C. Adams$^{1,2}$, David S. Graff$^2$, 
and Douglas O. Richstone$^2$} 
 
\affil{$^1$Physics Department, University of Michigan, Ann Arbor, MI 48109}

\affil{$^2$Astronomy Department, University of Michigan, Ann Arbor, MI 48109}

\begin{abstract} 
We construct a model for the formation of black holes within galactic
bulges. The initial state is a slowly rotating isothermal sphere,
characterized by effective transport speed $\aeff$ and rotation rate
$\Omega$. The black hole mass is determined when the centrifugal
radius of the collapse flow exceeds the capture radius of the central
black hole. This model reproduces the observed correlation between
black hole masses and galactic velocity dispersions, $\mbh \approx
10^8 M_\odot (\sigma/200 \kms)^4$, where $\sigma = \sqrt{2} \aeff$. 
This model also predicts the ratio $\mrat$ of black hole mass to host 
mass: $\mrat$ $\approx$ 0.004 $(\sigma/200 \kms)$. 
\end{abstract}


\section{INTRODUCTION} 

Two groups have recently reported an observed relationship between a 
galaxy's velocity dispersion $\sigma$ and the mass $\mbh$ of its central 
(supermassive) black hole. This correlation can be written in the form 
\be 
\mbh = M_0 \, (\sigma/200 \kms)^\gamma \, ,  
\label{eq:observed} 
\ee
where the two groups find $\gamma$ = 3.75 (Gebhardt et al. 2000) and
$\gamma$ = 4.7 (Farrarese \& Merritt 2000) and where the leading mass
coefficient is $M_0 \approx 1.2 - 1.3 \times 10^8$ $M_\odot$.  Both
groups report little scatter about the relationship (the scatter in
$\mbh$ at fixed $\sigma$ is bounded by 0.30 dex) and find no evidence
for a subsidiary dependence on the Hubble type, profile type, or
environment. This relationship supersedes a previous one claimed
between the black hole mass and bulge luminosity (Richstone et
al. 1998; Magorrian et al. 1998; van der Marel 2000; Kormendy \&
Richstone 1998). Furthermore, this measured scaling law poses a clear
challenge for galaxy formation theories, which must ultimately account
for this relationship.

Several theories imply a relationship between the central black hole
mass and the galactic velocity dispersion. For example, a
semi-analytic model of merger-driven starbursts with black hole
accretion (Haehnelt \& Kauffman 2000; Kauffman \& Haehnelt 2000)
provides a correlation of the form [\ref{eq:observed}].  Several
models are based on black hole accretion influencing star formation
and gas dynamics in the host galaxy; this feedback can occur through
ionization, mechanical work, and heating (e.g., Ciotti \& Ostriker
1997, 2000; Blandford 1999; Silk \& Rees 1998).  The model of
Blandford (1999) gives $\mbh < \eta \, \sigma^{5/3}$ whereas Silk \&
Rees (1998) predict $\mbh \propto \sigma^5$. Finally, the accretion of
collisional dark matter indicates the scaling relation $\mbh \propto
\sigma^{4 - 4.5}$ (Ostriker 2000).

In this letter, we present a new theory of the $\mbh - \sigma$
relation, based on an idealized model for the collapse of the inner
part of protogalaxies. The simplest variant of this theory is
described in terms of particle dynamics in \S 2. The role of gas
dynamics is explored in \S 3. We discuss the implications of this
model and identify future issues in \S 4.

\section{THE BASIC BALLISTIC MODEL} 

In this section, we examine the collapse of the inner part of a region
that will form the bulge of a galaxy. The calculation starts at the
time of maximum expansion for the main body of the bulge.  For the
sake of definiteness, we assume the following:
[1] The dark matter and baryons are unsegregated. 
[2] The mass in this region is distributed like a singular isothermal
sphere even though it is not in virial equilibrium. The initial density 
and mass distributions thus take the form 
\be 
\rho(r) = { \aeff^2 \over 2 \pi G r^2 } \qquad {\rm and} \qquad 
M (r) = {2 \aeff^2 \over G} \, r  \, . 
\label{eq:initial} 
\ee
The transport speed $\aeff$ that specifies the initial conditions is 
related to the isotropic velocity dispersion $\sigma$ according to 
$\sigma = \sqrt{2} \aeff$ (see below). 
[3] This region is slowly rotating like a solid body (e.g., due to
tidal torques) at a well-defined frequency $\Omega$.  Both dark matter
particles and parcels of baryons that are initially located at radius
$r_\infty$ thus have initial angular momentum $j = r_\infty^2 \Omega$
$\sin^2 \theta$, where $\theta$ is the polar angle in spherical 
coordinates. 
[4] The center of this region contains a black hole, which may have a
tiny mass at the start. This initial ``seed'' black hole could form by
the collapse of the densest (central) part of the perturbation or
could be primordial.

Particles in the main body of the initial distribution will fall
toward the center.  Because the dynamical time scales monotonically
increase with radius, infalling shells do not cross. The mass
contained inside a given spherical shell, which marks a particle's
location, does not change as the particle falls inward. As a result,
orbital energy is conserved and is given (classically) by 
\be
E = \half v_r^2 + \half {j^2 \over r^2} - {GM \over r} \, . 
\label{eq:energy} 
\ee 
We consider orbits which fall a long way toward the center of the
galaxy.  Idealizing these trajectories as zero energy orbits permits
the use of equation [\ref{eq:energy}] to determine their pericenters
$p$. For particles falling within the equatorial plane ($\theta =
\pi/2$), the pericenter can be written in the form 
\be
p = {j^2 \over 2GM} = {r_\infty^4 \Omega^2 \over 2GM} 
= {(GM)^3 \Omega^2 \over 2^5 \aeff^8} \, ,
\label{eq:peri} 
\ee 
where $r_\infty$ is the starting radius of an infalling particle and 
where we have used $M = M(r)$ as a label for $r_\infty$ in the final 
equality. 

If this pericenter $p$ is sufficiently small, ballistic particles will
pass inside the horizon of the black hole and be captured.  As mass
accumulates in the black hole, its horizon scale grows accordingly.
The pericenter of particles in ballistic orbits, falling from our
assumed mass distribution, increases as $p \propto r_\infty^3 \sim
M^3$.  In the earliest stages of the collapse, all of the falling
material is thus captured by the black hole.  Later, this growth
mechanism cuts off sharply when the black hole mass reaches a critical
point defined by equating the pericenter $p$ (for $\theta = \pi/2$
orbits) to the capture radius of the black hole. In Schwarzschild
geometry, particles coming in from infinity on zero energy orbits are
captured by the black hole if $p < 4 \rsw$ (Misner, Thorne, \& Wheeler
1973), where $\rsw = 2 GM/c^2$ is the Schwarzschild radius. The
condition $p=4 \rsw$ thus defines the critical mass scale $M_C$ where
direct accretion is compromised, i.e., 
\be
M_C \equiv { 16 \aeff^4 \over G c \Omega} \, =
{ 4 \sigma^4 \over G c \Omega} \, .  
\label {eq:mcrit} 
\ee 
In this scenario, the critical mass scale $M_C$ determines the
observed black hole mass $\mbh$ (note that equation [\ref{eq:mcrit}]
displays the correct scaling with $\sigma$).  Most of the baryonic
material not captured by the black hole during this early collapse
phase eventually forms stars in the galactic bulge. Dark matter with
low angular momentum is captured into the black hole along with the
baryons; dark matter with high angular momentum ($p > 4 \rsw$) passes
right through the galactic plane and forms an extended structure. 

We now use observations to specify the appropriate values of $\Omega$
and $\aeff$. In dissipationless collapse, the scale length of the mass
distribution drops by 1/2 from maximum expansion, and the ``flat
rotation curve conspiracy'' suggests that dissipation doesn't alter
$\sigma$ further.  This argument implies that the observed velocity
dispersion $\sigma$ is related to the initial transport speed $\aeff$
of the protogalactic material through the relation $\sigma^2 = 2
\aeff^2$.  To specify $\Omega$, we use the fundamental plane, which
provides a well defined relationship (for normal ellipticals and
bulges) between the half-light radii of galactic bulges and the
corresponding velocity dispersions (see Binney \& Merrifield 1998).
For $\sigma = 200\kms$, the effective radius $R$ of a mean surface
brightness elliptical on the fundamental plane is about 3.5 \kpc. With
this scale $R$ as the outer boundary of the forming bulge, the angular
momentum of a ballistic particle cannot exceed the angular momentum of
a circular orbit at this radius, i.e., $R^2 \Omega$ = $\sigma R$ or
$\Omega = \sigma/R$.  Evaluating this result for $\sigma = 200 \kms$,
we obtain a fiducial value of $\Omega = 5.8 \times 10^{-2} \myr^{-1}$
= $1.8 \times 10^{-15}$ rad s$^{-1}$. With the values of $\Omega$ and
$\sigma$ now specified, we use equation [\ref{eq:mcrit}] to find the
desired $\mbh - \sigma$ relation 
\be
\mbh \approx 10^8 \Msun (\sigma/200\kms)^4 \, , 
\label{eq:msig} 
\end{equation}
where we have written the result in terms of $\sigma$ rather than
$\aeff$. This relation is in good agreement with the observed
correlations (equation [\ref{eq:observed}]), as shown in Figure 1. 

\section{GAS DYNAMICS AND OTHER EFFECTS} 

In this section, we include gas dynamics in our model for black hole
formation during the collapse of galactic bulges. Because these
structures become gravitationally bound, we expect the proto-bulge to
collapse as a whole from an initial state (described here by equation
[\ref{eq:initial}]). To obtain a mathematical description of this
collapse, we consider the flow that produces a galactic bulge to be a
scaled up version of the collapse flows that have been studied
previously for star formation (Shu 1977; Terebey, Shu, \& Cassen
1984); this approach should thus capture the basic essence of the
collapse problem.  The collapse of the initial state (with density
distribution [\ref{eq:initial}]) proceeds from inside-out and the
central portion of the flow approaches a ballistic (pressure-free)
form: Dark matter always exhibits pressure-free behavior.  Even for
infalling gas, however, the inner limit of the collapse flow
approaches pressure-free conditions. For collapse over sufficiently
long time intervals, stars can form as parcels of gas fall inward and
the resulting infalling stars are manifestly ballistic.  The time
scale for individual star formation events is $\tstar \sim10^5$ years
(e.g., Adams \& Fatuzzo 1996), whereas the time scale $\tblg$ for the
entire bulge structure to form is much longer ($\tblg$ $\sim$ 25 -- 50
Myr).  We thus expect most of the stars to form while the overall
collapse of the bulge is still taking place.

For a given gravitational potential, we find the orbital solutions for
stars (or gas parcels or dark matter) falling towards the galactic
center. In this initial calculation, the inner solution is derived
using the gravitational potential of a point source. This form is only
used in the innermost regime of the collapse flow where the potential
is dominated by the forming black hole. As a result, this orbital
solution is valid for the range of length scales $\rsw \le r \ll 
r_\infty$. 
\footnote{Notice that at late times, long after black hole formation 
is complete, the potential is no longer close to a point potential 
and this solution loses its validity; in addition, relativistic 
corrections become important as $r\to\rsw$.} 
Since this potential is spherically symmetric, angular
momentum is conserved and the motion is confined to a plane described
by the coordinates $(r, \phi)$; the radius $r$ is the same in both the
plane and the original spherical coordinates. The angular coordinate
$\phi$ in the plane is related to the angle in spherical coordinates
by the relation $\cos \phi$ = $\cos \theta$ / $\cos \theta_0$, where
$\theta_0$ is the angle of the asymptotically radial streamline (see
below).  For zero energy orbits, the equations of motion imply a cubic
orbit solution, 
\be
1 - {\mu \over \mu_0} = (1 - \mu_0^2) \, 
{j_\infty^2 \over G M r} \, , 
\label{eq:orbit} 
\ee
where $j_\infty$ is the specific angular momentum of particles
currently arriving at the galactic center along the equatorial plane.
Here, the trajectory that is currently passing through the position
($r$, $\mu \equiv \cos \theta$) initially made the angle $\theta_0$
with respect to the rotation axis (where $\mu_0$ = $\cos \theta_0$). 
As in star formation theory, we define the centrifugal radius $R_C$  
$\equiv$ $j_\infty^2/GM$ ($=2p$), which represents the radius of a
circular orbit with angular momentum $j_\infty$ (for $\theta = \pi/2$). 
Our assumption of uniform initial rotation at rate $\Omega$ implies
that $j_\infty = \Omega r_\infty^2$, where $r_\infty$ is the starting
radius of the material that is arriving at the center at a given time.
To evaluate the radii $R_C$ and $r_\infty$, we invert the mass
distribution of the initial state (equation [\ref{eq:initial}]) to
find $r_\infty$=$GM/2\aeff^2$ and $R_C$ = $\Omega^2 G^3 M^3/16\aeff^8$. 

With an isothermal profile as the initial state, the collapse solution
indicates that the flow exhibits a well defined mass infall rate $\dot
M$ = $m_0 \aeff^3 /G$, where $m_0 \approx 0.975$ (Shu 1977).  This
infall rate is constant in time and we can measure the time since the
collapse began by the total mass $M$ that has fallen to the galactic
center. At early times, all of the mass falling to the center is
incorporated into the central black hole. At later times, the mass
supply is abruptly shut off by conservation of angular momentum. In
this setting, the mass infall rate is quite large, $\dot M$ $\approx$
650 $M_\odot$ yr$^{-1}$ (for $\sigma$ = 200 \kms and $\sigma^2 = 2
\aeff^2$). The time scale $\tbh$ to form a typical supermassive black
hole (with mass $\mbh \sim 10^8$ $M_\odot$) is thus about $\tbh \sim
10^5$ yr, comparable to the time scale $\tstar$ for individual stars
to form.  The time scale to form the entire bulge is much longer,
about $\tblg \sim 25 - 50$ Myr, comparable to the crossing time
$\tcross$ = $R/\aeff$.

In this collapse flow, streamlines entering the central region do not
cross each other. As long as the (bulge) infall time is longer than
the time scale for individual star formation events, the core regions
that produce stars will not interact. However, this scenario has an
initial transient phase ($\sim 10^5$ yr) in which the collapse of the
bulge structure takes place faster than individual stars form. If
stars are already condensing out of the collapse flow during this
initial phase, they can interact and merge upon entering the central
region. The resulting merger activity can lead to the initial
production of the central black hole, which then gains additional 
mass as the collapse proceeds. 

Given the orbital solution (equation [\ref{eq:orbit}]), we can find
the velocity fields for the collapse flow. The density distribution
$\rho(r,\theta)$ of the infalling material can then be obtained by
applying conservation of mass along a streamtube (Terebey, Shu, \&
Cassen 1984) and can be written in the form 
\be 
\rho(r,\theta) = { {\dot M} \over 4 \pi |v_r| r^2 }
{d \mu_0 \over d \mu} \, . 
\label{eq:density} 
\ee
The properties of the collapsing structure determine the orbit
equation [\ref{eq:orbit}], which in turn determines the form of
$d\mu_0/d\mu$ and the radial velocity $v_r$. The density field 
is thus completely specified (analytically, but implicitly). 

This gas dynamical version of the model defines the same critical mass
scale for central black holes as the ballistic model presented in \S
2.  In the earliest stages of collapse, incoming material falls to
small radii $r \ll \rsw$, where $\rsw \equiv 2 GM/c^2$ and $M$ is the
total mass $M={\dot M}t$ that has fallen thus far.  In this early
stage, the black hole mass $\mbh$ = $M$.  As the collapse develops,
incoming material originates from ever larger radii and carries a
commensurate increase in specific angular momentum. The centrifugal
barrier of the collapse flow grows with time.  The black hole mass is
determined when the centrifugal radius exceeds the capture radius of a
black hole in Schwarzschild geometry.  This condition takes the form
$R_C > \alpha \rsw$, where $\alpha = 8$ and is determined by particle
orbits in a Schwarzschild metric. This condition, $R_C > \alpha \rsw$,
leads to the same critical mass scale $M_C$ defined in equation
[\ref{eq:mcrit}].  Although $M_C$ determines the black hole mass
$\mbh$, two additional effects conspire to make the final black hole
mass somewhat larger:

[1] Even after the centrifugal barrier grows larger than the capture
radius, the black hole continues to gain mass from infalling
streamlines oriented along the rotational poles of the system. The
fraction of the infalling material that lands at such small radii is a
rapidly decreasing function of time. As a result, this effect makes
the black hole mass larger by only a modest factor $\fa$.  The mass
infall rate ${\dot M}_{\rm bh}$ for material falling directly onto the
black hole itself is given by 
\be
{\dot M}_{\rm bh} = \int_{-1}^1 d\mu \, 2\pi (\alpha \rsw)^2 
|v_r| \rho(\alpha \rsw, \mu) \, , 
\label{eq:addmass} 
\ee
where $\mu = \cos \theta$. Using equation [\ref{eq:density}] to
specify the density, we evaluate the integral to obtain a differential
equation for the time evolution of the black hole mass. Solving the
resulting differential equation (Adams et al. 2001), we find that the
black hole mass increases by a factor $\fa$ $\approx$ 1.35 due to
direct infall.

[2] Gas that falls to the midplane of the system can collect into a
disk structure surrounding the nascent black hole. The presence of the
disk is consistent with the current theoretical ideas about AGNs and
the jets they produce. In order to retain the desired scaling law
$\mbh \sim \sigma^4$, however, the total mass added to the black hole
through disk accretion must be less than (or comparable to) the
original mass scale $M_C$.  We can estimate the maximum amount of mass
that can be added to the black hole through the disk by using the
constraint that disk accretion cannot operate faster than the orbit
time at the outer disk edge (which is determined by $R_C$). For
reasonable assumptions, this maximum mass scale in the limit of
efficient disk accretion is about 10 $M_C$ (Adams et al. 2001).

This model also predicts a mass scale $M_B$ for the bulge itself. 
If the initial protobulge structure is rotating at angular velocity 
$\Omega$, then only material within a length scale $R = \aeff/\Omega$
can collapse to form the bulge. Material at larger radii, $r > R$,
is already rotationally supported and will not fall inwards.  The
length scale $R$ thus defines an effective outer boundary to the
collapsing region that forms the bulge. This boundary $R$, in turn,
defines a mass scale for the bulge, i.e., 
\be
M_B = {2 \aeff^3 \over G \Omega} \, 
\approx 2.4 \times 10^{10} M_\odot 
(\sigma / 200 \kms)^3 \, \, . 
\label{eq:mbulge} 
\ee
Both the bulge mass $M_B$ and the black hole mass $\mbh \approx M_C$
have the same dependence on the rotation rate $\Omega$. We can divide
out the rotation rate and find a robust estimate for the mass fraction 
$\mrat$, i.e., 
\be
\mrat \equiv {\mbh \over M_B} = 
\sqrt{32} \, {\sigma \over c} \approx 0.0038 \, 
(\sigma/200 \kms) \, . 
\label{eq:fraction} 
\ee
This mass fraction $\mrat$ is roughly comparable to the observed 
ratio of black hole masses to bulge masses in host galaxies (e.g., 
Richstone et al. 1998; Magorrian et al. 1998), although the data 
show appreciable scatter (see Figure 2). 

\section{CONCLUSION} 

In this paper, we have presented a simple model to describe the
collapse flow that produces galactic bulges and the supermassive black
holes living at their centers. The initial (pre-collapse) state is a
slowly rotating isothermal sphere, characterized by an effective sound
speed $\aeff$ and an angular velocity $\Omega$. These parameters
$(\aeff, \Omega)$ thus represent the specification of the initial
conditions. The velocity dispersion of the final stellar system is
comparable to the initial sound speed and we make the identification
$\sigma \approx \sqrt{2} \aeff$. In developing this basic picture, 
we find the following results: 

[1] The black hole mass $\mbh$ is determined by the condition that
the centrifugal radius exceeds the capture radius of a Schwarzschild
black hole. This requirement leads to the scaling law $\mbh = M_0
(\sigma/200 \kms)^4$, which is consistent with observations both in
its dependence on velocity dispersion $\sigma$ and for the mass scale
($M_0 \approx 10^8 \Msun$) of the leading coeffecient (see equation 
[\ref{eq:msig}] and Figure 1).

[2] The bulge mass is determined by the outer boundary of the
collapsing region -- material at initial radii $r > R$ is rotationally
supported and cannot collapse. This condition leads to the scaling law
$M_B \propto \sigma^3$ (equation [\ref{eq:mbulge}]) and predicts the
ratio $\mrat$ of black hole mass to bulge mass (equation 
[\ref{eq:fraction}]). This mass fraction scales weakly with velocity
dispersion and has a typical value $\mrat \approx 0.004$, similar to
observed mass ratios (Figure 2). 

Elliptical galaxies (and bulges) are described by four parameters:
effective radius $R_e$, luminosity $L$, velocity dispersion $\sigma$,
and central black hole mass $\mbh$. For this discussion, we substitute
the total mass $M_B$ for the luminosity (for a given dark matter
fraction, we thus assume that the transformation $M_B \to L$ is
determined by known, but perhaps complicated, stellar physics). We
need to make the connection between these four basic quantities
$(\sigma, R_e, M_B, \mbh)$ and our model, which is described by only
two variables: the transport speed $\aeff$ and the rotation rate $\Omega$. 
Our model implies the following transformation between initial 
conditions and final system properties:
[1] $\sigma = \sqrt{2} \aeff$, 
[2] $R_e = \aeff/ (2 \Omega)$, 
[3] $M_B = 2 \aeff^3 / (G \Omega)$, 
[4] $\mbh = 16 \fa \aeff^4 / (c G \Omega)$, and 
[5] $\mrat = 8 \fa \aeff/c$, where $\fa \approx 1.35$. 
The fifth relation is not independent, but because the rotation rate 
cancels out, it represents a robust prediction. The implications of 
these five scaling laws must be tested further against observations. 

In this letter, we have presented a working collapse model for black
hole and bulge formation. In future work, a number of issues should be
addressed, including relativistic corrections to the infall solutions,
additional black hole mass contributions from disk accretion, and
feedback from the central black hole on the collapse. In order to
successfully reproduce the observed correlations, our model uses
initial conditions with a particular profile of specific angular
momentum (determined by the initial $\rho(r)$ and $\Omega$); we thus
need to explore more general initial conditions and make the
connection between the required initial conditions and earlier stages
of galaxy formation. In addition, galactic bulges are likely to
experience merger events (White \& Rees 1978) which lead to modest
increases in the overall velocity dispersion (White 1979); we must
determine how the black holes produced through this model evolve in
the context of such galactic mergers. In any case, the properties of
these supermassive black holes -- and their connection to their host
galaxies -- will ultimately provide a vital diagnostic for the galaxy
formation process.

\acknowledgments
 
We would like to thank Gus Evrard, Luis Ho, Greg Laughlin, Manasse
Mbonye, and Scott Tremaine for useful discussions. This work was
supported by funding from the University of Michigan, bridging support
from NASA Grant No. 5-2869, the NASA Long Term Space Astrophysics
program, and Space Telescope Science Institute.

\vskip 0.5truein 
\centerline{\bf FIGURE CAPTIONS} 
\bigskip 

\noindent 
Figure 1. The correlation between black hole mass $\mbh$ and velocity
dispersion $\sigma$ of the host galaxy. The data points (adapted from
Gebhardt et al. 2000) represent the observed correlation for
ellipticals (circles), S0 galaxies (squares), and spirals (triangles).
The solid curve shows the theoretical result of this paper (using 
equation [\ref{eq:msig}] and the 35\% correction from equation
[\ref{eq:addmass}]). The dashed and dotted curves show the
observational fits advocated by Gebhardt et al. (2000) and 
Farrarese \& Merritt (2000), respectively. 

\bigskip 
\noindent 
Figure 2. The ratio $\mrat$ of black hole mass to host mass plotted 
as a function of the velocity dispersion $\sigma$ of the host galaxy. 
The solid curve shows the prediction of this paper.  The data points
(adapted from Gebhardt et al. 2000) exhibit considerable scatter, but
their mean value is in reasonable agreement with theoretical
expectations; the various symbols represent ellipticals (circles), 
S0 galaxies (squares), and spirals (triangles). 

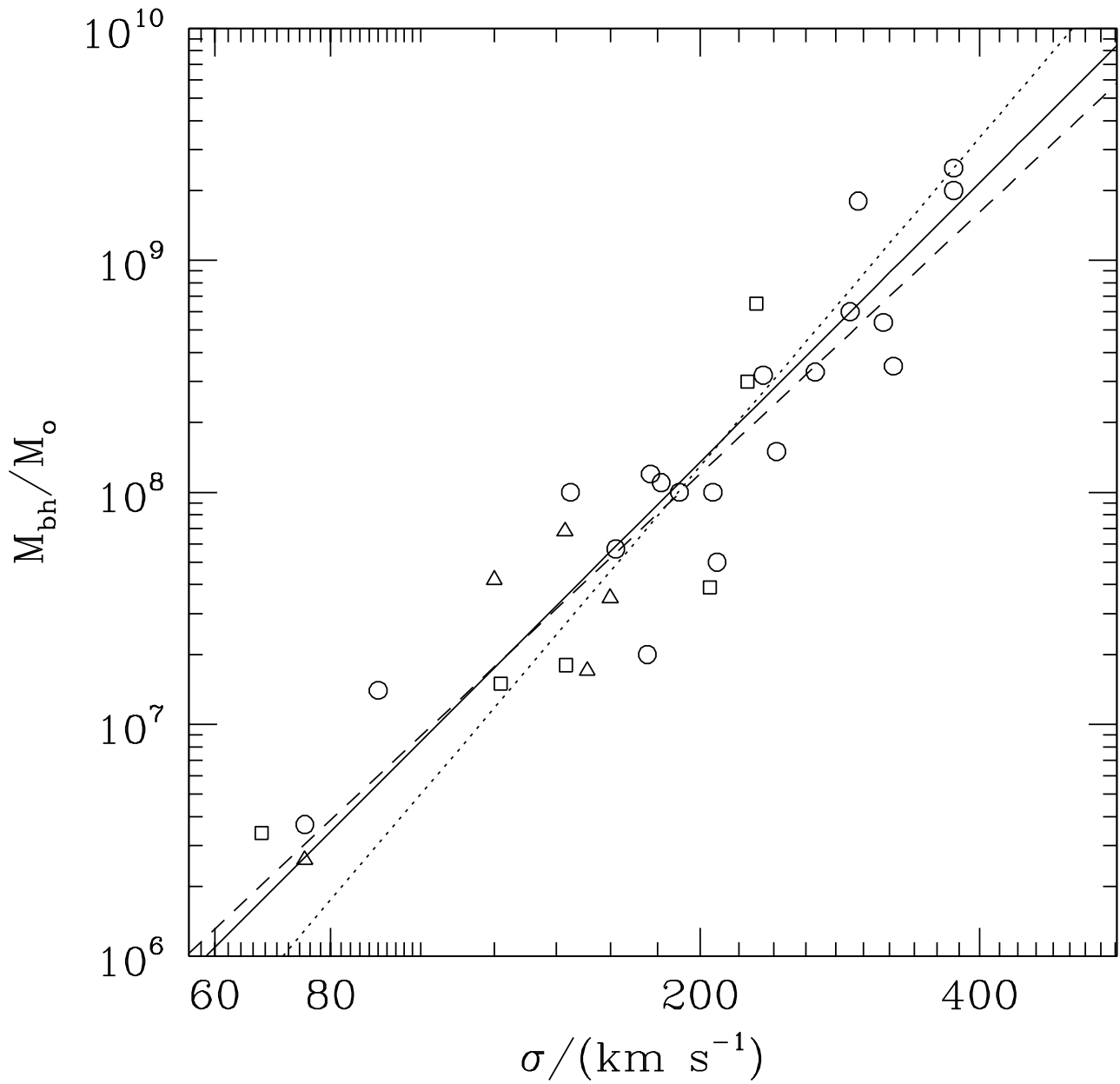
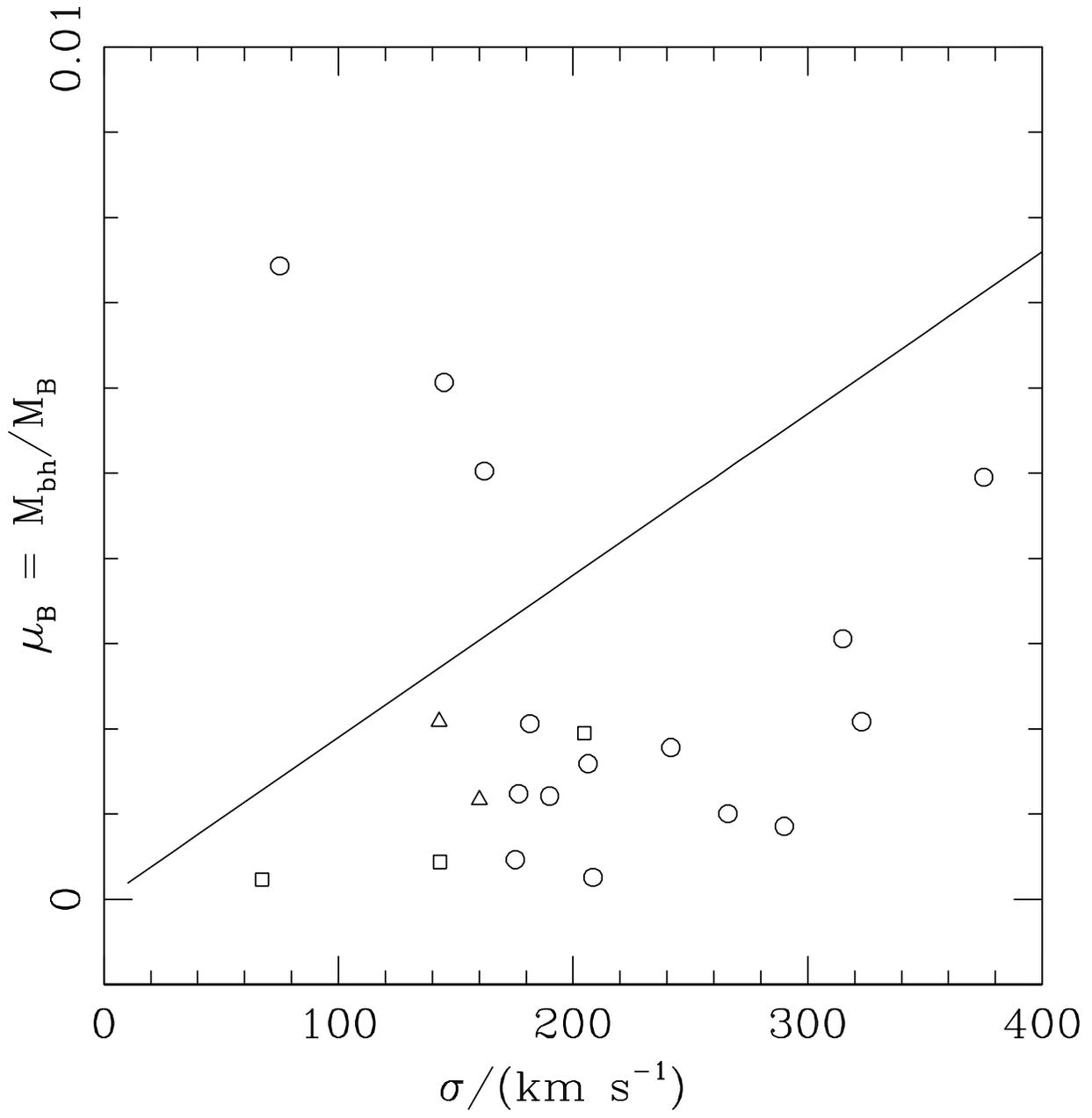
\end{document}